\PassOptionsToPackage{dvipsnames,table}{xcolor}
\documentclass[aps,pre,10pt,twocolumn,superscriptaddress,longbibliography,floatfix,showkeys]{revtex4-2}

\usepackage{xcolor}
\usepackage{amsmath,amssymb,amsthm}
\newtheorem{defn}{Definition}
\usepackage{graphicx}
\usepackage{tikz}
\usetikzlibrary{matrix}
\definecolor{coopColor}{RGB}{235, 235, 235}   
\definecolor{defColor}{RGB}{60, 60, 60}       

\newcommand{\C}{\node[fill=coopColor, minimum size=5.0mm, anchor=center,
                       text=black] {\textbf{C}};}
\newcommand{\D}{\node[fill=defColor, minimum size=5.0mm, anchor=center,
                       text=white] {\textbf{D}};}

\newcommand{\Cfn}[1]{\node[fill=coopColor, minimum size=5.0mm, anchor=center,
                            text=black, draw=black, line width=1.5pt]
                            {\textbf{C}\textsubscript{#1}};}
\newcommand{\Dfn}[1]{\node[fill=defColor, minimum size=5.0mm, anchor=center,
                            text=white, draw=white, line width=1.5pt]
                            {\textbf{D}\textsubscript{#1}};}
\tikzset{
    gamegrid/.style={
        matrix of nodes,
        nodes in empty cells,
        column sep=-\pgflinewidth,
        row sep=-\pgflinewidth,
        nodes={draw=gray!30, minimum size=5.0mm, anchor=center}
    }
}

\begin{document}
\emergencystretch=1em
\hbadness=10000
\hfuzz=6pt
\vfuzz=2pt

\title{From Local Payoffs to Global Instabilities:
A Spectral Cartography of Spatiotemporal Chaos in Canonical
$2\times2$ Evolutionary Games}

\author{Ozgur Aydogmus}
\email{ozgur.aydogmus@asbu.edu.tr}
\affiliation{Department of Economics, Social Sciences University of
             Ankara, Ankara, T\"{u}rkiye}

\date{\today}

\begin{abstract}
We develop a motif-based framework for spatiotemporal chaos in spatial evolutionary games and use it to map the dynamical phase diagram in the payoff plane. Using Boolean linearization of the imitate-the-best rule, we derive analytical instability thresholds for local motifs including invaders, cooperative pairs, stripe interfaces, and cooperative cores. These thresholds are obtained from payoff balance at contested motif interfaces and recover classical invasion thresholds of spatial evolutionary games, which emerge here as boundaries of the chaotic phase. Combining the Derrida slope with the asymptotic Hamming distance, we obtain a four-region cartography: ordered, transient-chaotic, sustained-chaotic, and subcritical-chaotic dynamics. The phase diagram is organized by density-dependent motif selection: different initial cooperator densities activate different instability mechanisms, yet a small set of motif-instability lines consistently bounds the sustained-chaos region across densities. This cartography reveals a subcritical chaotic phase (Derrida slope $s<1$ but asymptotic Hamming distance $d_\infty>0$), where infinitesimal perturbations decay while finite-amplitude perturbations sustain chaos. The motif-based framework is anchored by an exact benchmark: for homogeneous backgrounds, the Boolean Jacobian yields an exact correspondence between the Derrida slope and spectral radius, linking damage spreading to deterministic instability.\end{abstract}


\maketitle

\section{Introduction}
\label{sec:intro}

Nowak and May~\cite{nowakmay} observed that the deterministic
``imitate-the-best'' update rule generates chaotic dynamics in the
spatial Prisoner's Dilemma game. Simple local interactions give rise
to complex and persistent spatial patterns, allowing cooperators and
defectors to coexist through the constant formation and breakup of
clusters.
This result sparked broad interest in how spatial structure acts as a
mechanism for sustaining cooperation. A large body of work has
since examined spatial evolutionary games on lattices and complex
networks, focusing on how interaction topology shapes
asymptotic cooperator
frequencies~\cite{wang2015evolutionary,szabo2007evolutionary,
hauert2002effects,szabo1998evolutionary}; for
comprehensive reviews of spatial social dilemmas from a
statistical-physics perspective, see
Refs.~\cite{perc2017statistical,jusup2022social}. Yet while spatial
pattern formation and phase transitions between strategy
configurations are by now well reviewed, the chaotic dynamical
richness identified by Nowak and May---transient structures,
sensitivity to initial conditions, and irregular spatiotemporal
behavior---has received comparatively less attention. At the population level, the mean-field counterpart of these models (e.g.
the replicator equations) is well understood
dynamically~\cite{hofbauer1998evolutionary}, but it smooths away the spatial heterogeneity that drives the phenomena of
interest here.

Many dynamical models of evolutionary games arise as large-population
limits of microscopic update rules, giving rise either to well-mixed
ODEs or to spatially extended lattice and continuum
formulations capable of pattern formation and
instability~\cite{hofbauer1998evolutionary,benaim2003deterministic,
hwang2013deterministic,aydogmus2017preservation,aydogmus2018discovering}.
These models, including discrete ones, represent real-world
collective-action problems such as social cooperation, public goods
provisioning, and epidemiological
spread~\cite{Gao2024,Zhang2025,Traulsen2023}.
Understanding when such models develop instability
and emergent complexity requires methods that operate directly at the
level of the microscopic update rule.

The present paper addresses this gap through a unifying perspective:
the onset of macroscopic damage spreading in spatial evolutionary
games is controlled by the loss of stability of characteristic local
spatial motifs. Each instability threshold is set by
a simple condition: the two strategies competing at the motif's
interface earn equal payoffs. Different motifs generate different
instability mechanisms and therefore different phase boundaries in
the payoff plane. The resulting motif-instability structure organizes
the observed chaotic regions and explains their dependence on the
initial cooperator density.

Two complementary analytical traditions address stability in
dynamical systems. The first is Fourier stability analysis, which
linearizes the governing dynamics and diagnoses spatial instabilities
through the growth of plane-wave perturbations; it is the standard
tool for continuum systems and coupled-map
lattices~\cite{murraymathematical,aydogmus2018discovering,kaneko1989pattern}.
The second is damage spreading, introduced by Derrida and
Pomeau~\cite{derrida1986random}, which quantifies chaotic sensitivity
in discrete symbolic systems through the probabilistic growth of a
small perturbation between two initially near-identical replicas.
The present paper connects these two perspectives in
the setting of spatial evolutionary games. The linear stability of
small characteristic spatial patterns---the \emph{motifs} defined in
Sec.~\ref{sec:framework}---controls the onset of macroscopic damage
spreading, and hence the location of the chaotic phase boundaries in
the payoff plane.

Alfaro and Sanju\'{a}n~\cite{Alfaro2024} recently established the
normalized Hamming distance as a reliable numerical indicator of
spatial chaos in evolutionary games, saturating within the chaotic
regimes originally observed by Nowak and May and vanishing otherwise. While their work provides a robust numerical characterization, an analytical framework linking the onset of spatial chaos to the microscopic update dynamics has remained unavailable.

A fundamental obstacle to linking microscopic update rules with
macroscopic instability is the discrete and non-differentiable nature
of cellular-automaton dynamics. Because each update
selects the maximum among the finitely many payoffs in an
agent's neighborhood, rather than
applying a smooth transformation of the state, conventional
Jacobian-based linearization is unavailable. We overcome this by
working in the sparse-damage limit, where the initially perturbed
sites are sufficiently rare that they evolve independently. In this
regime the Boolean
derivative~\cite{vichniac1990boolean,luque2000lyapunov} yields an
exact linear operator governing first-order perturbation propagation,
providing a spectral description of instability in an otherwise
non-smooth system.

For a homogeneous reference background, translational
invariance renders the Boolean Jacobian a block-circulant convolution
operator, exactly diagonalized by the discrete Fourier transform; its
spectral radius $\Lambda=\max_{\mathbf{k}}|\lambda(\mathbf{k})|$
coincides exactly with the Derrida slope ($s=\Lambda$).
In this exactly solvable case, damage spreading and
spectral instability coincide. The Derrida contours mapped in
Sec.~\ref{qc}, however, are obtained from random initial conditions,
where the lattice forms a mosaic of locally ordered motifs (uniform
regions, stripes, and cooperative cores). Within this motif-based
effective-medium picture, the onset of damage spreading is associated
with the first statistically prevalent motif whose linearized dynamics
becomes marginally unstable ($\Lambda_M=1$). Our computations further
reveal a subcritical regime ($s<1$, $d_\infty>0$) in which
finite-amplitude perturbations sustain chaos despite linear stability.

The critical condition $s\approx1$ can be expressed
analytically through payoff-balance relations, one for each spatial
motif. This criterion analytically predicts the boundaries of the
spatiotemporal chaos regions originally observed through numerical
simulation by Nowak and May~\cite{nowakmay}\footnote{Nowak and May
employed a von Neumann neighborhood for payoff calculation and a Moore
neighborhood for strategy comparison. Here a von Neumann neighborhood
is used for both; although quantitative thresholds differ, the
analysis extends to any translationally invariant neighborhood.} and
supplies a unifying spectral interpretation of the discrete
cluster-invasion thresholds derived in later
literature~\cite{hauert2002effects,szabo2007evolutionary,
hauert2001fundamental}.
The connection between these thresholds and chaotic phase boundaries
had not previously been established.

Combining the Derrida slope with the asymptotic Hamming distance
$d_\infty$~\cite{Alfaro2024}, we construct a spectral cartography of
the full $(u,v)$ payoff plane distinguishing ordered behavior,
transient chaos, sustained chaos, and subcritical chaos across
multiple initial cooperator densities. The main contributions of this work are therefore:

\begin{enumerate}
\item \textbf{Motif-controlled chaotic cartography.}
The four-regime phase diagram is organized by the full set of
motif-instability lines, with the initial cooperator density selecting
which motifs are active. Within this organization a small, fixed
subset (notably Lines~III, V, and~VIII) bounds the
\emph{sustained}-chaos region at every density studied, so the
topology of the phase diagram is largely invariant with cooperator
density rather than fragmenting into density-specific cases.

\item \textbf{A subcritical chaotic phase.}
We identify and map a regime with $s<1$ yet $d_\infty>0$: infinitesimal
perturbations decay, but finite-amplitude perturbations sustain chaos,
so the ordered state is metastable.
This is a dynamical phase distinct from linear instability, not a
reinterpretation of any invasion threshold.

\item \textbf{Analytical prediction of phase boundaries.}
Closed-form motif marginal-stability conditions---obtained from
payoff balance at each motif's contested interface---reproduce the
known cluster-invasion thresholds and are identified, for the first
time, as the boundaries of the chaotic phase.

\item \textbf{A Boolean spectral framework.}
We connect probabilistic damage spreading, the
spectral stability of the Boolean Jacobian, and motif geometry. For
homogeneous backgrounds the connection is exact: the Derrida slope
equals the spectral radius, $s=\Lambda$.
\end{enumerate}

The paper is organized as follows.
Section~\ref{sec:model} introduces the model.
Section~\ref{qc} presents the numerical chaos measures.
Section~\ref{sec:framework} develops the analytical framework.
Section~\ref{sec:hamming} classifies the dynamical regimes.
Section~\ref{sec:conclusio} concludes.

\section{Canonical $2\times2$ Games and the Model}
\label{sec:model}

We study the deterministic cellular automaton introduced by Nowak and
May~\cite{nowakmay,nowakmay2}, in which \emph{canonical $2\times2$
games} are played on a square lattice of size $L\times L$.
Each site $s_{ij}$ is occupied by an individual who adopts one of two
strategies, cooperator ($C$) or defector ($D$), and plays against
their four von Neumann neighbors.
Payoffs are determined by the two-parameter matrix for canonical
$2\times2$ games~\cite{aydogmus2020does}:
\begin{equation}
\begin{pmatrix} R & S \\ T & P \end{pmatrix}
=
\begin{pmatrix} 1 & v \\ 1+u & 0 \end{pmatrix},
\end{equation}
where $u = T-R$ measures the \textit{temptation to defect} and
$v = S-P$ the \textit{risk of cooperation}.
Restricting to $-1\le u,v\le1$ is sufficient to
capture the four canonical game families~\cite{minireview}, whose
defining regions partition the payoff plane as shown in
Fig.~\ref{fig:chaotic_boundaries_first}(c):
\begin{itemize}
    \item \textit{Prisoner's Dilemma} ($u>0$, $v<0$): defection
          strictly dominates cooperation;
    \item \textit{Snowdrift / Hawk--Dove} ($u>0$, $v>0$): a weak
          dilemma in which cooperators and defectors can coexist;
    \item \textit{Stag Hunt} ($u<0$, $v<0$): a coordination dilemma
          in which mutual cooperation is optimal but fragile against
          strategic uncertainty;
    \item \textit{Harmony or Invisible Hand} ($u<0$, $v>0$):
          cooperation strictly dominates, so no social dilemma
          arises~\cite{bowles2003microeconomics,dawes1980}.
\end{itemize}
These outcomes are valid for well-mixed populations~\cite{aydogmus2020does}. Spatial structure can alter them substantially---sometimes
sustaining cooperation where it would collapse in a well-mixed
setting, and sometimes generating persistent chaotic dynamics. We study these spatial effects below.

\subsection*{Synchronous imitate-the-best rule}

At each discrete time step $t\in\mathbb{N}$, every individual
simultaneously computes their own payoff and that of each neighbor,
then adopts the strategy of whoever scored highest, retaining their
current strategy in case of a tie.
This synchronous ``imitate-the-best'' rule defines a nonlinear CA
whose linearized dynamics we analyze in Sec.~\ref{sec:framework}.
We focus on synchronous updates, whose discrete translational
invariance underlies the spectral analysis of
Sec.~\ref{sec:framework}; possible extensions to asynchronous and
probabilistic rules are discussed in Sec.~\ref{sec:conclusio}.

\subsection*{Payoff functions}

For a focal individual with $n_C\in\{0,1,2,3,4\}$ cooperative
neighbors, the payoffs under each strategy are:
\begin{align}
\Pi_C(n_C) &= n_C + (4-n_C)\,v, \label{eq:payoff_C}\\
\Pi_D(n_C) &= n_C\,(1+u). \label{eq:payoff_D}
\end{align}
Under the imitate-the-best rule, each agent compares its own payoff with those of its four neighbors and adopts the strategy of the highest scorer. At the contested interface of a spatial motif, the decisive comparison is therefore between a cooperator--defector pair, with each agent's payoff evaluated at its \emph{own} cooperative-neighbor count: $\Pi_C(n_C)$ for the cooperator versus $\Pi_D(n_C')$ for the defector, where $n_C$ and $n_C'$ generally differ. Equating these two payoffs yields the payoff-balance condition of the motif (Table~\ref{tab:final_unified}).

\section{Quantifying Chaos: Hamming Distance, Damage Spreading,
         and the Derrida Slope}
\label{qc}

Most studies of the spatial games track asymptotic cooperator
frequencies~\cite{hauert2002effects,hauert2001fundamental}.
We instead ask whether the dynamics is ordered or chaotic,
quantifying sensitivity to initial conditions through damage
spreading~\cite{derrida1986random,derrida1,derrida2}.

\begin{figure*}[t]
\centering
\includegraphics[width=\textwidth]{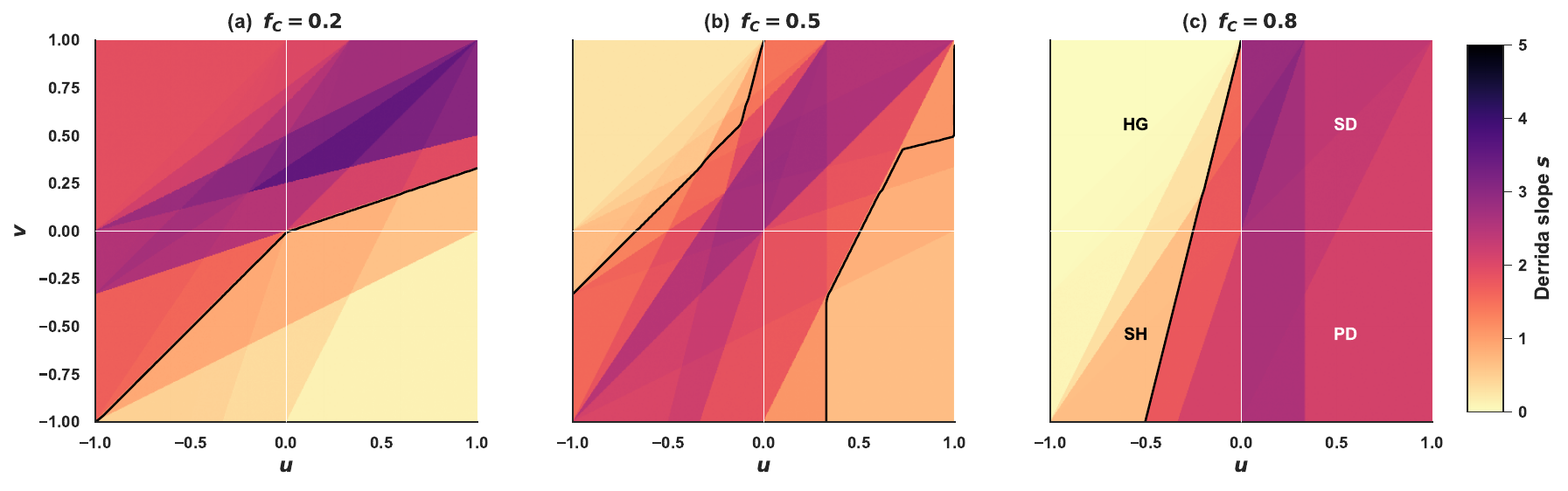}
\caption{\textbf{The chaotic phase boundary shifts systematically with
initial cooperator density, implying distinct dominant instability
mechanisms at each density.} Each panel shows the Derrida slope $s(u,v)$ for a fixed initial cooperator frequency $f_C$; darker colors indicate $s>1$ (chaotic phase) and lighter colors $s<1$ (ordered phase). Black contours mark the empirical chaotic boundaries $s\approx1$. The shape and position of these contours shift substantially with $f_C$ (see e.g. ``chaotic corridor'' appearing only at $f_C=0.5$) indicating that the dominant instability mechanism depends on the initial spatial composition of the population. Section~\ref{sec:framework} identifies which motif (and its corresponding spectral label) is responsible at each boundary. The canonical game regions---Harmony (HG), Stag Hunt (SH), Prisoner's Dilemma (PD), and Snowdrift (SD)---are annotated in the rightmost panel for reference.}
\label{fig:chaotic_boundaries_first}
\end{figure*}

The damage-spreading protocol works as follows.
A reference configuration $A_0$ is initialized randomly for a given
$(u,v)$ pair, and a perturbed copy $B_0$ is obtained by flipping the
strategies at a small number of randomly chosen sites, introducing
initial damage $d_0$.
Both replicas are then evolved for one time step under the identical
imitate-the-best rule $\mathcal{F}$: $A_1=\mathcal{F}(A_0)$,
$B_1=\mathcal{F}(B_0)$.
The resulting damage $d_1$ is measured by the normalized Hamming
distance.

\begin{defn}[Hamming Distance]\label{def:hamming}
For two configurations $A,B$ on the $L\times L$ lattice, the
normalized Hamming distance is the fraction of sites at which their
strategies differ:
\begin{equation}
d(A,B)=\frac{1}{L^2}\sum_{n,m=1}^{L}
\mathbf{1}\{A(n,m)\neq B(n,m)\},
\end{equation}
where $\mathbf{1}\{\cdot\}$ is the indicator function.
We write $d_t:=d(A_t,B_t)$ for the damage at time $t$.
\end{defn}

Repeating this protocol over many initial-condition pairs gives the
Derrida map, whose slope at the origin quantifies the \emph{average
one-step amplification} of an infinitesimal perturbation.

\begin{defn}[Derrida Map and Slope]\label{derrida}
The Derrida map $g(d_0)=\mathbb{E}[d_1\,|\,d(A_0,B_0)=d_0]$ is the
expected post-update damage averaged over an ensemble of
configuration pairs with the same initial damage $d_0$.
The Derrida slope is its derivative at the origin:
\begin{equation}
s=\left.\frac{dg(d_0)}{dd_0}\right|_{d_0\to0}.
\end{equation}
\end{defn}

The Derrida slope $s$ is the discrete-system analogue of the largest
Lyapunov multiplier: it measures the average per-step amplification of
an infinitesimal perturbation in a symbolic system like a
CA~\cite{luque2000lyapunov,vispoel2024damage}. Thus $s>1$ signals
exponential growth of infinitesimal damage (chaotic phase), $s<1$
contraction (ordered phase), and $s=1$ the critical boundary.
Because the dynamics is Boolean, no tangent space
exists and a Lyapunov exponent cannot be defined in the usual way; the
Derrida slope is the standard substitute for such symbolic systems,
defined through the ensemble average of Definition~\ref{derrida}. The
connection between $s$ and the spectral properties of the linearized
rule is the subject of Sec.~\ref{sec:framework}. CA can exhibit mixed
behavior in which ordered and chaotic regions coexist in the same
rule~\cite{wolfram1984universality,langton1990computation}; here we
find that the location and extent of the chaotic region depend not
only on the payoff parameters $(u,v)$ but also on the initial
cooperator density $f_C(0)$, which sets the prevalence of the local
motifs that drive instability.

We scan the $(u,v)$ plane on $50\times50$ lattices\footnote{Simulations on $100\times100$ and $200\times200$ lattices yield indistinguishable phase diagrams, confirming that the boundaries are not finite-size artifacts.} for $f_C(0)\in\{0.2,0.5,0.8\}$, averaging $s$ over 30 independent initial conditions at each parameter point. Figure~\ref{fig:chaotic_boundaries_first} summarizes the results and
reveals two findings that motivate the framework developed
in Sec.~\ref{sec:framework}. First, the empirical
chaotic phase boundaries (black contours, $s\approx1$) are sharp and
approximately linear in the $(u,v)$ plane, suggesting that each
boundary is governed by a single dominant instability mechanism.
Second, both the location and the extent of the chaotic phase depend
strongly on the initial cooperator frequency $f_C(0)$: a ``chaotic
corridor'' spanning much of the SD and SH regions appears only at
$f_C(0)=0.5$, while high and low cooperator densities yield different
chaotic regions with qualitatively different boundary geometries. The
central question is therefore: which spatial motif becomes unstable at
each boundary, and why does the dominant motif change with $f_C(0)$?
Section~\ref{sec:framework} answers this question.

\section{A Unified Analytical Framework for Stability Boundaries}
\label{sec:framework}

The numerical results of Sec.~\ref{qc} establish that the Derrida
slope $s$ transitions sharply across approximately linear boundaries
in the $(u,v)$ plane, and that the location of those boundaries
depends on the initial cooperator density.
The Derrida slope, however, measures only the \emph{average} growth
of a spatially random perturbation: it tells us \emph{that} a
transition occurs, but not \emph{which feature of the local update
rule} is responsible.

To identify the instability mechanism, we shift from random
perturbations to structured ones.
Specifically, for an ordered reference background we ask how a
plane-wave perturbation of wavevector $\mathbf{k}$ is amplified under
a single application of the linearized update rule---a similar
question asked of PDEs via linear stability analysis and of
coupled-map lattices via Fourier
decomposition~\cite{cross1993,kaneko1989pattern}.
The key insight is that particular instability
mechanisms can be associated with characteristic spatial motifs
(homogeneous domains, stripes, and cooperative cores) together with
their interface perturbations such as stripe kinks. Each ordered
background is naturally associated with a Fourier or Bloch
decomposition, and the marginal-stability condition of that motif
translates directly into a payoff-balance relation between the two
competing strategies at the motif's characteristic local geometry.
The chaotic phase boundaries observed numerically are then the lines
in the $(u,v)$ plane at which one of these balance conditions is
first satisfied---that is, at which the corresponding motif becomes
marginally unstable.

The section is organized as follows.
We first derive the Fourier-mode linearization and establish the link
between the Derrida slope and the spectral radius
(Sec.~\ref{subsec:formalism}).
We then derive each motif's payoff-balance line
(Sec.~\ref{subsec:derivations}).
Finally, we show which motif dominates at each initial cooperator
density and compare the predicted lines with the numerical boundaries
(Sec.~\ref{subsec:derrida}).

\subsection{General Formalism: Linearization and Modal Decomposition}
\label{subsec:formalism}

\paragraph*{1.~Linearization of the update rule.}
Let $\delta_{n,m}(t)\in\{0,1\}$ denote the perturbation field: $\delta_{n,m}=1$
if site $(n,m)$ differs between the two replicas
(Definition~\ref{def:hamming}). When the initial damage is sparse, damaged
sites are separated by several lattice spacings and evolve independently.
Perturbation propagation is then governed to leading order by the Boolean
Jacobian of the update
rule~\cite{vichniac1990boolean,luque2000lyapunov,vispoel2024damage},
\begin{equation}
\delta_{n,m}(t+1)=
\sum_{(r,s)\in\mathcal{N}_0} J_{n,m}(r,s)\,\delta_{n+r,m+s}(t),
\label{eq:linearized_general}
\end{equation}
where $\mathcal{N}_0$ contains the focal site and its von Neumann neighbors,
and the Boolean derivative
$J_{n,m}(r,s)=\partial s_{n,m}(t+1)/\partial s_{n+r,m+s}(t)\in\{0,1\}$
equals $1$ if flipping the state at offset $(r,s)$ changes the next state of
$(n,m)$. Because it is set by the local payoff ordering, $J_{n,m}(r,s)$
generally varies from site to site; its structure depends on the reference
background.

We adopt a strict von Neumann neighborhood for both payoff accumulation and
imitation, whereas Nowak and May~\cite{nowakmay} mixed von Neumann and Moore
neighborhoods. Since the analysis below relies only on discrete translational
invariance, it extends to Moore or other symmetric neighborhoods with the
quantitative thresholds shifting accordingly.

\paragraph*{2.~Homogeneous backgrounds: dispersion relation and the exact
identity $s=\Lambda$.}
When every site holds the same strategy, the payoff ordering---and hence the
Boolean derivative---is identical at every site, and the Jacobian reduces to a
scalar convolution,
\begin{equation}
\delta_{n,m}(t+1) = J_0\,\delta_{n,m}(t)
  + J_*\!\!\sum_{(n',m')\in\mathcal{N}_{n,m}}\!\!\delta_{n',m'}(t).
\label{eq:linearized_rule}
\end{equation}
Because the background is uniform and the rule isotropic, the sensitivity
$J_*$ is the same for all four neighbors; this is the only setting in which
that is guaranteed. Substituting the plane-wave ansatz
$\delta_{n,m}(t)=\tilde\delta_{\mathbf{k}}(t)\,e^{i(k_x n+k_y m)}$, the focal
site contributes $J_0$ and the four neighbors contribute
$2J_*(\cos k_x+\cos k_y)$, so the modes decouple:
\begin{eqnarray}
& &\tilde\delta_{\mathbf{k}}(t+1) =
\lambda(\mathbf{k})\,\tilde\delta_{\mathbf{k}}(t),\nonumber\\
& &\lambda(\mathbf{k}) = J_0 + 2J_*[\cos(k_x)+\cos(k_y)].
\label{eq:lambda_general}
\end{eqnarray}
The linearized dynamics is stable if and only if the spectral radius
$\Lambda=\max_{\mathbf{k}}|\lambda(\mathbf{k})|\le1$.

For this background the Derrida slope coincides exactly with the spectral
radius,
\begin{equation}
s = \Lambda,
\label{eq:derrida_spectral}
\end{equation}
even though the former is an average over all modes and the latter a maximum.
The two agree because the Boolean coefficients are idempotent
($J_0^2=J_0$, $J_*^2=J_*$): the Brillouin-zone average of
$|\lambda(\mathbf{k})|^2$ equals $J_0^2+4J_*^2=J_0+4J_*$, which is also
$\max_{\mathbf{k}}|\lambda(\mathbf{k})|$
(Appendix~\ref{app:idempotency}). The identity~\eqref{eq:derrida_spectral}
is exact and anchors the framework: in the solvable homogeneous case, damage
spreading and spectral instability are the same quantity.

\paragraph*{3.~Periodic and random backgrounds.}
For a background with a nontrivial unit cell---a stripe or a cooperative
core---the Boolean derivative differs between site classes within the cell.
The Jacobian is then circulant at the level of unit cells (block-circulant)
and is diagonalized by a Bloch--Fourier transform, with marginal stability
given by $\rho(\widehat{J}(\mathbf{k}))=1$, where
$\widehat{J}(\mathbf{k})$ is a finite Bloch symbol
(Appendix~\ref{app:bloch}). 

\begin{figure*}[tbp]
\centering
\includegraphics[width=\textwidth]{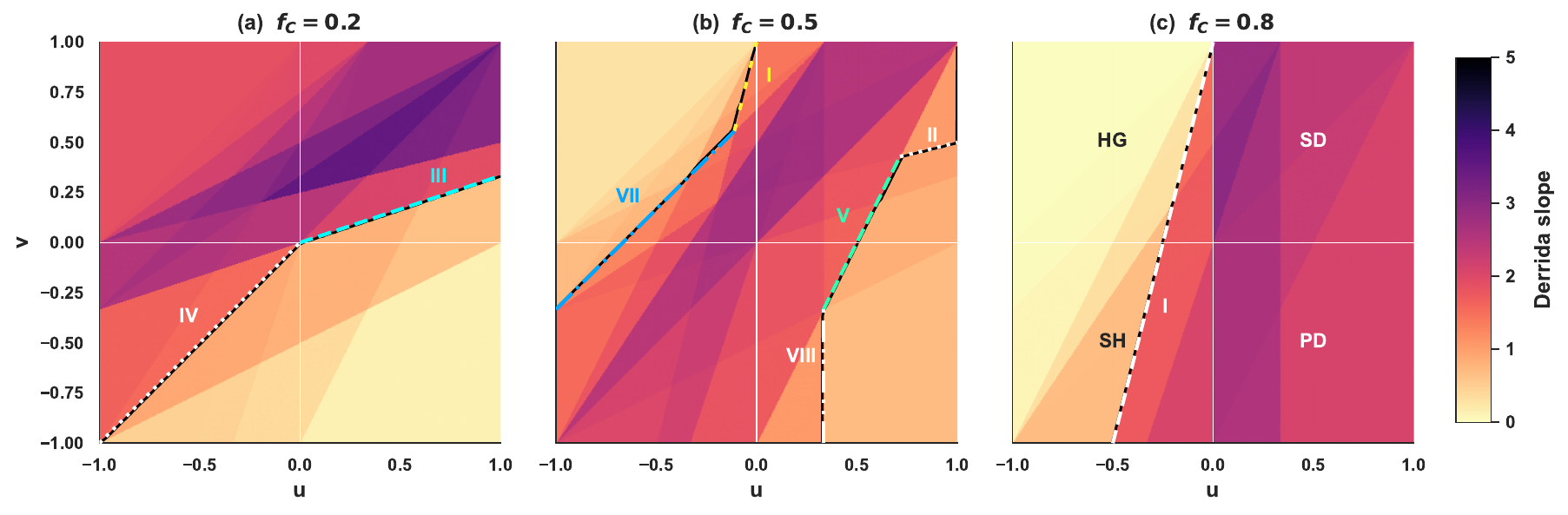}
\caption{\textbf{The analytically derived instability lines
(Table~\ref{tab:final_unified}) quantitatively predict the numerical
chaos boundaries, with each boundary segment attributable to a single
dominant motif.}
Black contours show the empirical Derrida boundary $s\approx1$ from
Fig.~\ref{fig:chaotic_boundaries_first}; colored dashed lines are the
analytical predictions from Table~\ref{tab:final_unified}.
Panel~(a) ($f_C=0.2$): the chaotic region is bounded below by the
stripe equilibrium line~IV ($v=u$) and the nucleation
line~III ($v=u/3$).
Panel~(b) ($f_C=0.5$): the upper boundary of the ``chaotic corridor''
follows lines~VII and~I; the lower boundary is a
composite of lines~V, II, and~VIII.
Panel~(c) ($f_C=0.8$): dominated by the single-defector invasion
line~I ($v=4u+1$).
The shift in dominant line across panels reflects how $f_C$ controls
which local geometric motif is statistically prevalent in the
reference configuration.}
\label{fig:chaotic_boundary_lines}
\end{figure*}

For a random lattice at density $f_C$ no
translational invariance survives. We treat such a lattice as a mosaic of
locally ordered patches, each homogeneous or periodic over a few lattice
spacings, so that the local spectral radius $\Lambda_M$ of a prevalent motif
$M$ governs damage growth within its patches. In the thermodynamic limit the
onset of global damage spreading is then set by the first statistically
prevalent motif to reach marginal stability,
\begin{equation}
s=1 \;\Longleftrightarrow\; \Lambda_M=1
\quad\text{for the first prevalent motif } M,
\label{eq:threshold}
\end{equation}
an effective-medium equivalence established in
Appendix~\ref{app:threshold}. It applies in
the limit $d_0\to0$ probed by the Derrida slope, the regime in which linear
stability analysis is valid~\cite{strogatz2015nonlinear}, and the resulting boundaries
agree with the numerical contours of
Fig.~\ref{fig:chaotic_boundary_lines} throughout the parameter range studied.

\paragraph*{4.~From spectral instability to payoff geometry.}
In all non-homogeneous cases we obtain the thresholds not by diagonalizing
$\widehat{J}(\mathbf{k})$ but directly from payoff balance. At the interface
of each characteristic motif there is a pair of competing agents---one
cooperator and one defector---whose relative payoffs determine whether the
motif grows or shrinks. Marginal stability is reached when the two attain
equal payoff, so that neither strategy can invade the other's territory. This
\emph{payoff-balance condition} is linear in $(u,v)$ and is the analytical
origin of the nearly linear boundaries in
Fig.~\ref{fig:chaotic_boundaries_first}. The procedure for each motif is:
(i)~identify the ordered reference pattern it represents, (ii)~locate the
competing pair at its contested interface, (iii)~set their payoffs equal using
Eqs.~\eqref{eq:payoff_C}--\eqref{eq:payoff_D}, and (iv)~solve for the line in
the $(u,v)$ plane. Only two structural properties of the dynamics are used:
translational invariance (block-circulant Jacobian) and Boolean, idempotent
switching coefficients; the game payoffs enter only through step~(iii). The
framework therefore applies to any CA sharing these properties.

\subsection{Derivation of Critical Boundaries from Motif Geometry
            and Marginal Stability}
\label{subsec:derivations}

\begin{figure}[tbp]
\centering
\begin{minipage}{\columnwidth}
\centering
{\small \textbf{Long-wavelength mode
$\mathbf{k}\!\to\!(0,0)$}}\par\smallskip
\begin{minipage}[b]{0.3\columnwidth}
\centering
\begin{tikzpicture}
\matrix[gamegrid]{
  \C & \C & \C & \C \\
  \C & \C & \Cfn{3} & \C \\
  \C & \C & \Dfn{4} & \C \\
  \C & \C & \C & \C \\
};
\end{tikzpicture}
\par\smallskip{\footnotesize (a) Line I: D in all-C}
\end{minipage}
\hfill
\begin{minipage}[b]{0.3\columnwidth}
\centering
\begin{tikzpicture}
\matrix[gamegrid]{
  \D & \D & \D & \D \\
  \D & \D & \Dfn{1} & \D \\
  \D & \D & \Cfn{0} & \D \\
  \D & \D & \D & \D \\
};
\end{tikzpicture}
\par\smallskip{\footnotesize (b) Line II: C in all-D}
\end{minipage}
\hfill
\begin{minipage}[b]{0.3\columnwidth}
\centering
\begin{tikzpicture}
\matrix[gamegrid]{
  \D & \D & \D & \D \\
  \D & \Cfn{1} & \Dfn{1} & \D \\
  \D & \C & \D & \D \\
  \D & \D & \D & \D \\
};
\end{tikzpicture}
\par\smallskip{\footnotesize (c) Line III: pair vs D}
\end{minipage}
\end{minipage}

\vspace{0.45cm}

\begin{minipage}{\columnwidth}
\centering
{\small \textbf{Stripe modes
$\mathbf{k}=(0,\pi),(\pi,0)$}}\par\smallskip
\begin{minipage}[b]{0.3\columnwidth}
\centering
\begin{tikzpicture}
\matrix[gamegrid]{
  \D & \C & \D & \C \\
  \D & \C & \D & \C \\
  \D & \Cfn{2} & \Dfn{2} & \C \\
  \D & \C & \D & \C \\
};
\end{tikzpicture}
\par\smallskip{\footnotesize (d) Line IV: self-dual}
\end{minipage}
\hfill
\begin{minipage}[b]{0.3\columnwidth}
\centering
\begin{tikzpicture}
\matrix[gamegrid]{
  \D & \C & \D & \C \\
  \C & \Cfn{3} & \Dfn{2} & \C \\
  \D & \C & \D & \C \\
  \D & \C & \D & \C \\
};
\end{tikzpicture}
\par\smallskip{\footnotesize (e) Line V: Cooperator kink}
\end{minipage}
\hfill
\begin{minipage}[b]{0.3\columnwidth}
\centering
\begin{tikzpicture}
\matrix[gamegrid]{
  \C & \D & \C & \D \\
  \D & \Dfn{1} & \Cfn{2} & \D \\
  \C & \D & \C & \D \\
  \C & \D & \C & \D \\
};
\end{tikzpicture}
\par\smallskip{\footnotesize (f) Line VI: Defector kink}
\end{minipage}
\end{minipage}

\vspace{0.45cm}

\begin{minipage}{\columnwidth}
\centering
\begin{minipage}[b]{0.45\columnwidth}
\centering
{\small \textbf{Oblique
$\mathbf{k}=(\pi,\pi/2)$}}\par\smallskip
\begin{tikzpicture}
\matrix[gamegrid]{
  \C & \D & \C & \D \\
  \C & \Dfn{3} & \Cfn{1} & \D \\
  \D & \C & \D & \C \\
  \D & \C & \D & \C \\
};
\end{tikzpicture}
\par\smallskip{\footnotesize (g) Line VII: oblique interface}
\end{minipage}
\hfill
\begin{minipage}[b]{0.45\columnwidth}
\centering
{\small \textbf{Core $\mathbf{k}=(0,\pi/3)$}}\par\smallskip
\begin{tikzpicture}
\matrix[gamegrid]{
  \C & \C & \C & \D \\
  \C & \Cfn{4} & \C & \D \\
  \C & \C & \Dfn{3} & \D \\
  \C & \C & \C & \D \\
};
\end{tikzpicture}
\par\smallskip{\footnotesize (h) Line VIII: core erosion}
\end{minipage}
\end{minipage}

\caption{\textbf{Each chaotic phase boundary traces to a single
microscopic competition event, fully determined by the payoff
functions.}
Each panel shows the local configuration (motif) associated with the
labeled sector; light cells are cooperators (C), dark cells are
defectors (D).
Cells with subscripted letters mark the critical competing pair whose
payoff balance defines the motif's marginal-stability condition;
subscripts give each agent's cooperative-neighbor count $n_C$.
Panels (a--c) show that the long-wavelength limit
$\mathbf{k}\to(0,0)$ captures localized invasion geometries (single
invader or two-site cluster); panels (d--f) probe domain-wall
stability and its transverse kink instabilities; panels (g--h)
correspond to higher-wavenumber instabilities that break the stripe
symmetry or erode cooperative cores.}
\label{fig:micro_grids}
\end{figure}
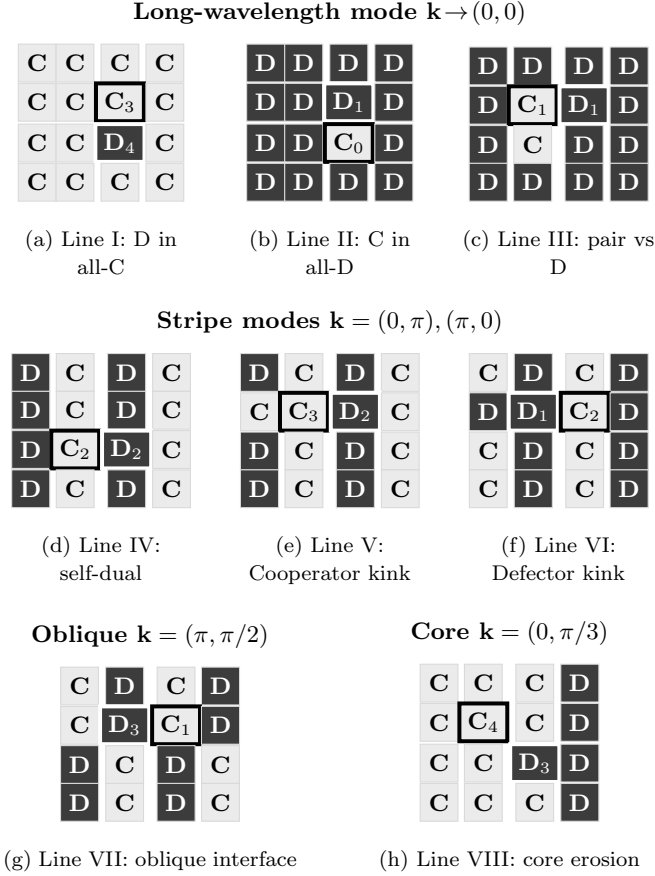

We now apply the procedure of
Sec.~\ref{subsec:formalism}, Step~4 to each characteristic motif:
identify the ordered reference pattern, locate the competing pair at
its contested interface, and set their payoffs equal to obtain the
marginal-stability line in the $(u,v)$ plane. The microscopic
geometries are illustrated in Fig.~\ref{fig:micro_grids} and all
results are collected in Table~\ref{tab:final_unified}.

Several of the payoff-balance conditions derived below have appeared
previously as cluster invasion criteria.
Hauert~\cite{hauert2001fundamental} showed that a $2\times2$
cooperator cluster persists when $2+2S>T$ and derived analogous
inequalities for $1\times1$ and $3\times3$ clusters on both
von~Neumann and Moore neighborhoods; Szab\'{o} and
F\'{a}th~\cite{szabo2007evolutionary} obtained similar thresholds
across the literature. These earlier results were obtained by direct
score comparison at specific cluster geometries, without connecting
those thresholds to spectral stability or chaotic dynamics. The
present framework reinterprets each invasion criterion as the
payoff-balance condition of a specific motif at marginal stability. To
our knowledge, this is the first work to place the known
cluster-invasion thresholds within a unified damage-spreading
framework and to relate them systematically to the observed chaotic
phase boundaries.

The wavevector assigned to each motif is a
representative label of the lowest-wavenumber periodic background
possessing the corresponding symmetry; it organizes the motifs but is
not an input to any computation. The thresholds themselves follow from
payoff balance at each motif's contested interface.

\paragraph*{1.~Long-Wavelength Sector $(\mathbf{k}\to(0,0))$.}

The conditions in this sector describe localized
nucleation events---a single invader, or the smallest cooperative
cluster, embedded in a homogeneous background. For the single-invader
cases (Lines~I and~II) the marginal-stability condition is read
directly from the Boolean derivative: by definition $J_*=1$ precisely
when flipping one site changes the focal update, so the threshold is
the locus where $J_*$ switches value---equivalently, where the invader
and the adjacent defender attain equal payoff. Line~III is the
smallest \emph{two-site} cluster (a cooperative pair, each member at
$n_C=1$ through mutual support); being a two-site object it is not a
single-site sensitivity but a cluster-nucleation threshold, obtained
from the same payoff-balance principle at the pair's contested
interface. None of the three requires a spectral computation.

We retain the label $\mathbf{k}\to(0,0)$ only for consistency with the
spectral framework: a localized nucleus is a Kronecker delta, hence
broadband in Fourier space, exciting all modes equally. Since
$J_0,J_*\ge0$, the homogeneous dispersion
$\lambda(\mathbf{k})=J_0+2J_*[\cos k_x+\cos k_y]$ is maximized at the
uniform mode, so the fastest-growing component of that broadband
perturbation is the $\mathbf{k}\to(0,0)$ sector. The label is therefore
an interpretation, not an input to the derivation, and not a claim
that the nucleus is a literal $\mathbf{k}=(0,0)$ eigenmode.

\begin{itemize}
    \item \emph{Defector invading all-$C$}
    (Fig.~\ref{fig:micro_grids}(a)):
    A single defector surrounded by cooperators has $n_C=4$, while
    its nearest cooperator neighbor has $n_C=3$.
    From Eqs.~\eqref{eq:payoff_C}--\eqref{eq:payoff_D},
    $\Pi_D(4)=4(1+u)$ and $\Pi_C(3)=3+v$.
    Setting these equal gives $\boxed{v=4u+1}$ (Line~I).

    \item \emph{Cooperator invading all-$D$}
    (Fig.~\ref{fig:micro_grids}(b)):
    A single cooperator surrounded by defectors has $n_C=0$; its
    nearest defector neighbor has $n_C=1$ (one cooperative neighbor
    from the invading cluster).
    Then $\Pi_C(0)=4v$ and $\Pi_D(1)=1+u$, giving
    $\boxed{v=(u+1)/4}$ (Line~II).

    \item \emph{Pair invasion / filament nucleation}
    (Fig.~\ref{fig:micro_grids}(c)):
    A two-site cooperative cluster embedded in a defector background:
    each cooperator in the pair has one cooperative neighbor and three
    defector neighbors ($n_C=1$), and competes with an adjacent
    defector that also has $n_C=1$.
    Then $\Pi_C(1)=1+3v$ and $\Pi_D(1)=1+u$, giving
    $\boxed{v=u/3}$ (Line~III).\footnote{The symmetric case---a
    defector pair invading an all-$C$ background---gives $v=3u$:
    each defector has $n_C=3$ and competes with a cooperator also
    having $n_C=3$, so $\Pi_C(3)=3+v$
    vs.\ $\Pi_D(3)=3(1+u)$.}
\end{itemize}

\paragraph*{2.~Stripe Modes $(\mathbf{k}=(0,\pi),(\pi,0))$.}

The stripe mode corresponds to alternating columns (or rows) of
cooperators and defectors.
In this striped background every agent has
exactly two cooperative neighbors and two defector neighbors
($n_C=2$), regardless of their own strategy.
This is the \emph{self-dual point} of the dynamics: cooperators and
defectors experience identical neighborhood compositions, so the
stripe is neutrally stable---neither strategy has a payoff advantage
at the interface.
The balance condition is simply
$\Pi_C(2)=\Pi_D(2)\Rightarrow\boxed{v=u}$ (Line~IV).
Since $\Pi_C(2)>\Pi_D(2)$ for $v>u$ and the reverse for $v<u$, the
$C$--$D$ domain wall favors cooperators above this line and defectors
below it---domain healing versus roughening in the language of the
equivalent ferromagnetic (coordination-game) kinetic-Ising
mapping~\cite{szabo2007evolutionary}.

Transverse displacements of the stripe front create
\emph{kinks}---sites where the local neighborhood deviates from the
ideal $n_C=2$ configuration:
\begin{itemize}
    \item \emph{Cooperator kink} (Fig.~\ref{fig:micro_grids}(e)):
    a cooperator protruding into defector territory acquires an extra
    cooperative neighbor ($n_C=3$) and competes with an adjacent
    defector still at $n_C=2$.
    $\Pi_C(3)=\Pi_D(2)\Rightarrow\boxed{v=2u-1}$ (Line~V).

    \item \emph{Defector kink} (Fig.~\ref{fig:micro_grids}(f)):
    a defector protruding into cooperator territory reduces the
    cooperative-neighbor count of an adjacent defector to $n_C=1$,
    which then competes with a cooperator still at $n_C=2$.
    $\Pi_C(2)=\Pi_D(1)\Rightarrow\boxed{v=(u-1)/2}$ (Line~VI).
\end{itemize}

\paragraph*{3.~Oblique Mode
$(\mathbf{k}=(\pi,\pi/2),(\pi/2,\pi))$.}

The oblique mode breaks the square-lattice symmetry
between horizontal and vertical directions, so its contested interface
pairs two agents with markedly different neighborhood compositions.
In the configuration of Fig.~\ref{fig:micro_grids}(g), a defector
has $n_C=3$ while the adjacent cooperator has only $n_C=1$.
$\Pi_C(1)=\Pi_D(3)\Rightarrow\boxed{v=u+2/3}$ (Line~VII).

\paragraph*{4.~Core Mode $(\mathbf{k}=(0,\pi/3),(\pi/3,0))$.}

When cooperative clusters are thick enough, interior
cooperators acquire four cooperative neighbors ($n_C=4$) and are
entirely shielded from direct interaction with defectors.
Instability of the cluster then originates not at the outermost
boundary but one layer in, where a ``pocket'' defector at $n_C=3$
competes with a fully shielded core cooperator at $n_C=4$
(Fig.~\ref{fig:micro_grids}(h)).
Since $\Pi_C(4)=4$ is independent of $v$, the balance condition
involves only $u$:
$\Pi_C(4)=\Pi_D(3)\Rightarrow\boxed{u=1/3}$ (Line~VIII).
This is the only vertical boundary line in
Table~\ref{tab:final_unified}, reflecting the fact that core erosion
is driven purely by the temptation to defect $u$ and is insensitive
to the risk of cooperation $v$.

\begin{table*}[tbp]
\centering
\caption{\textbf{The eight analytically derived instability lines
that constitute the chaotic phase boundaries in the $(u,v)$ payoff
plane.}
Each line is the locus at which the named motif transitions from
stable to unstable under the imitate-the-best rule.
The dominant active line at a given initial cooperator density $f_C$
determines the chaotic phase boundary observed numerically in
Fig.~\ref{fig:chaotic_boundary_lines}; which line dominates depends
on which local geometric motif is statistically prevalent at that
density (see Sec.~\ref{subsec:derrida}).}
\label{tab:final_unified}
\begin{ruledtabular}
\begin{tabular}{lllll}
Label & Critical line & Spectral label ($\mathbf{k}$)
      & Geometric configuration & Microscopic mechanism \\
\hline
I   & $v=4u+1$     & Long-wavelength $(0,0)$  & Single D in all-C
    & Defector invasion \\
II  & $v=(u+1)/4$  & Long-wavelength $(0,0)$  & Single C in all-D
    & Cooperator invasion \\
III & $v=u/3$      & Long-wavelength $(0,0)$  & 1D filament / pair
    & Nucleation threshold \\
IV  & $v=u$        & Stripe $(\pi,0)/(0,\pi)$ & Self-dual stripe
    & Domain-wall equilibrium \\
V   & $v=2u-1$     & Stripe $(\pi,0)/(0,\pi)$ & Cooperator kink
    & Transverse propagation \\
VI  & $v=(u-1)/2$  & Stripe $(\pi,0)/(0,\pi)$ & Defector kink
    & Domain roughening \\
VII & $v=u+2/3$    & Oblique $(\pi,\pi/2)$    & Oblique interface
    & Oblique instability \\
VIII& $u=1/3$      & Core $(0,\pi/3)$         & Pocket D vs core C
    & Core erosion \\
\end{tabular}
\end{ruledtabular}
\end{table*}

\subsection{Connecting Derrida Slopes with Motif Prevalence}
\label{subsec:derrida}

The analytical lines obtained above and summarized in
Table~\ref{tab:final_unified} predict eight candidate boundaries in
the $(u,v)$ plane. Which line is observed numerically depends on which
motif is most readily excited by the initial configuration---and that depends on the initial cooperator density $f_C(0)$.
A random lattice at density $f_C(0)$ predominantly
contains the local motifs that are statistically most likely at that
density, and the observed boundary is set by whichever of those motifs
first loses marginal stability. We treat three representative
densities in turn.

\paragraph{Low initial cooperation $(f_C(0)=0.2)$.}
When cooperators are rare, the lattice is predominantly defector
territory and the statistically prevalent motifs are small
cooperative clusters embedded in a defector background.
The relevant motifs are therefore those clusters themselves
(isolated pairs and single sites), governed by the long-wavelength
nucleation lines, together with the $C$--$D$ domain walls that bound
any cooperative cluster, governed by the self-dual stripe condition
(Line~IV).
In the SD region ($u>0$, $v>0$), the dominant
boundary is Line~III ($v=u/3$, pair nucleation): a two-site
cooperative cluster can sustain itself and grow only above this
threshold, so the chaotic phase is bounded below by this line.
In the SH region ($u<0$, $v<0$) the game is a coordination game with
two stable phases---all-$C$ and all-$D$---and Line~IV ($v=u$) is its
self-dual coexistence point. Following the kinetic-Ising
interpretation of Szab\'{o} and F\'{a}th~\cite{szabo2007evolutionary},
the $C$--$D$ domain wall carries a surface tension that changes sign
across this line: above it ($v>u$) cooperative domains have positive
surface tension and expand, while below it ($v<u$) they contract.
At low $f_C(0)$ the lattice is seeded in the all-$D$
basin with sparse cooperative nuclei, so Line~IV is the operative boundary because it determines whether those nuclei grow---nucleating the cooperative phase and escaping the all-$D$ basin---or are absorbed. Although labeled the self-dual \emph{stripe} condition in
Table~\ref{tab:final_unified}, Line~IV operates as the
coexistence condition of the bistable Stag-Hunt game at low $f_C(0)$, not as a
signature of prevalent stripes.
\begin{figure*}[tbp]
\centering
\includegraphics[width=\textwidth]{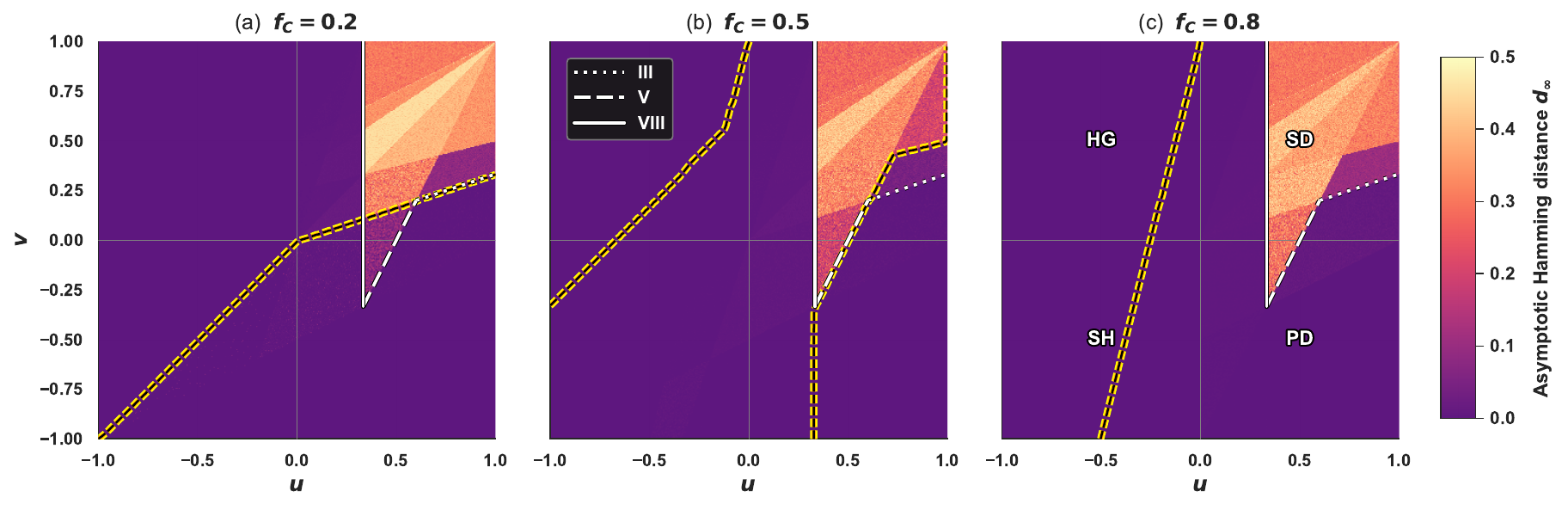}
\caption{\textbf{Two distinct chaos regimes coexist: transient chaos
(large) and sustained chaos (smaller interior region), with a
subcritical pocket where linear stability and persistent chaos
coexist.}
Each panel shows $d_\infty(u,v)$, the long-run fraction of sites that
differ between two initially close replicas, for initial cooperator
frequency $f_C(0)=0.2$ (a), $0.5$ (b), $0.8$ (c).
Color encodes $d_\infty$ on a perceptually uniform
sequential scale: dark purple indicates $d_\infty\approx0$
(eventual synchronization) and lighter colors indicate
$d_\infty>0$ (sustained chaos).
Yellow contours reproduce the empirical Derrida boundary $s\approx1$
from Fig.~\ref{fig:chaotic_boundaries_first}; white lines bound
the region of sustained chaos ($d_\infty>0$).
The gap between the yellow and white lines identifies
the \emph{transient chaos} regime: parameters where $s>1$ yet
$d_\infty\approx0$, meaning perturbations initially grow but are
eventually absorbed as the system organizes into stable geometric
domains.
The small regions inside the white lines but outside the yellow
contour---visible in panels (a) and (b)---are the
\emph{subcritical chaos} regime ($s<1$, $d_\infty>0$), discussed in
Sec.~\ref{sec:hamming}.}
\label{fig:hamming}
\end{figure*}
\paragraph{High initial cooperation $(f_C(0)=0.8)$.}
When the lattice is nearly all-cooperator, stability is governed by
the resilience of the cooperative background against a single
invading defector~\cite{hauert2002effects}.
For sufficiently large $u$, a lone defector surrounded by cooperators
achieves the highest local payoff ($\Pi_D(4)=4+4u$) and is
immediately imitated by all four neighbors, triggering a cascade of
defector
spreading~\cite{szabo2007evolutionary,hauert2001fundamental}.
This is precisely the Line~I scenario
(see Fig.~\ref{fig:micro_grids}(a)): the Derrida boundary is dominated
by $v=4u+1$, as confirmed by Fig.~\ref{fig:chaotic_boundary_lines}(c).
Because the reference configuration is nearly homogeneous at
$f_C(0)=0.8$, the Boolean Jacobian is close to its all-$C$ form and the
exact identity $s=\Lambda$ (Eq.~\eqref{eq:derrida_spectral}) holds to
good accuracy---consistent with the sharp boundary observed
numerically. Once Line~I is crossed, the Derrida slope exceeds unity and the system may enter the spatiotemporal chaos regime characterized by
``kaleidoscope'' patterns and traveling waves of
defectors~\cite{szabo2007evolutionary,hauert2001fundamental}.

\paragraph{Balanced initial conditions $(f_C(0)=0.5)$.}
At equal initial frequencies, cooperator and defector motifs coexist
in comparable proportions, and several motifs approach marginal
stability simultaneously, producing the composite ``chaotic
corridor'' visible in Fig.~\ref{fig:chaotic_boundary_lines}(b). The upper boundary of the corridor is defined by
whichever of Line~I ($v=4u+1$, single-defector invasion) and
Line~VII ($v=u+2/3$, oblique interface) is first crossed as $(u,v)$
moves into the chaotic region: these two lines govern
the shattering of large uniform strategy blocks into propagating
alternating structures.
The lower boundary is a composite of Line~V ($v=2u-1$, cooperator
kink), Line~II ($v=(u+1)/4$, cooperator invasion of all-$D$), and
Line~VIII ($u=1/3$, core erosion). Line~VIII is the only vertical boundary---a purely $u$-driven, core-erosion transition: for $u>1/3$ a pocket defector ($n_C=3$)
out-scores even a fully shielded core cooperator ($n_C=4$),
independently of $v$, so the core site adopts defection at the next
update and the shielded interior loses marginal stability. At
$f_C(0)=0.5$ there is sufficient cooperator material for clusters thick
enough to develop shielded $n_C=4$ cores, so this motif is prevalent
and its erosion threshold becomes an active boundary---unlike the
sparse, thin clusters at low density.

For balanced initial conditions, the reference configuration is maximally disordered, so the exact identity $s=\Lambda$---which holds only for a homogeneous background---does not apply. Nevertheless, the close agreement between the analytically derived boundaries and the numerical Derrida contours in Fig.~\ref{fig:chaotic_boundary_lines}(b) is consistent with the threshold equivalence (Eq.~\eqref{eq:threshold}): the Derrida slope crosses unity as the most statistically prevalent local motifs---cooperative pairs, stripes, kinks, and cores---successively lose marginal stability, in line with the effective-medium argument of Appendix~\ref{app:threshold}.

\section{Asymptotic Hamming Distance and Dynamical Regions}
\label{sec:hamming}

The Derrida slope $s$ is a local quantity: it measures the
infinitesimal, one-step growth of damage and therefore characterizes only the \emph{onset} of instability.
Whether that instability sustains itself over long times---or is
eventually absorbed by the system's attractor---is measured by the asymptotic Hamming distance $d_\infty$~\cite{Alfaro2024}.
A nonzero $d_\infty$ means that two initially close
replicas never synchronize: the system remains in a state of
perpetual dynamical activity.
By contrast, $d_\infty\approx0$ means the replicas eventually
converge, even if their trajectories diverged transiently.
Together, $s$ and $d_\infty$ allow us to distinguish four
qualitatively different dynamical regimes, which we now map across
the $(u,v)$ plane.
\begin{figure*}[t]
\centering
\includegraphics[width=\textwidth]{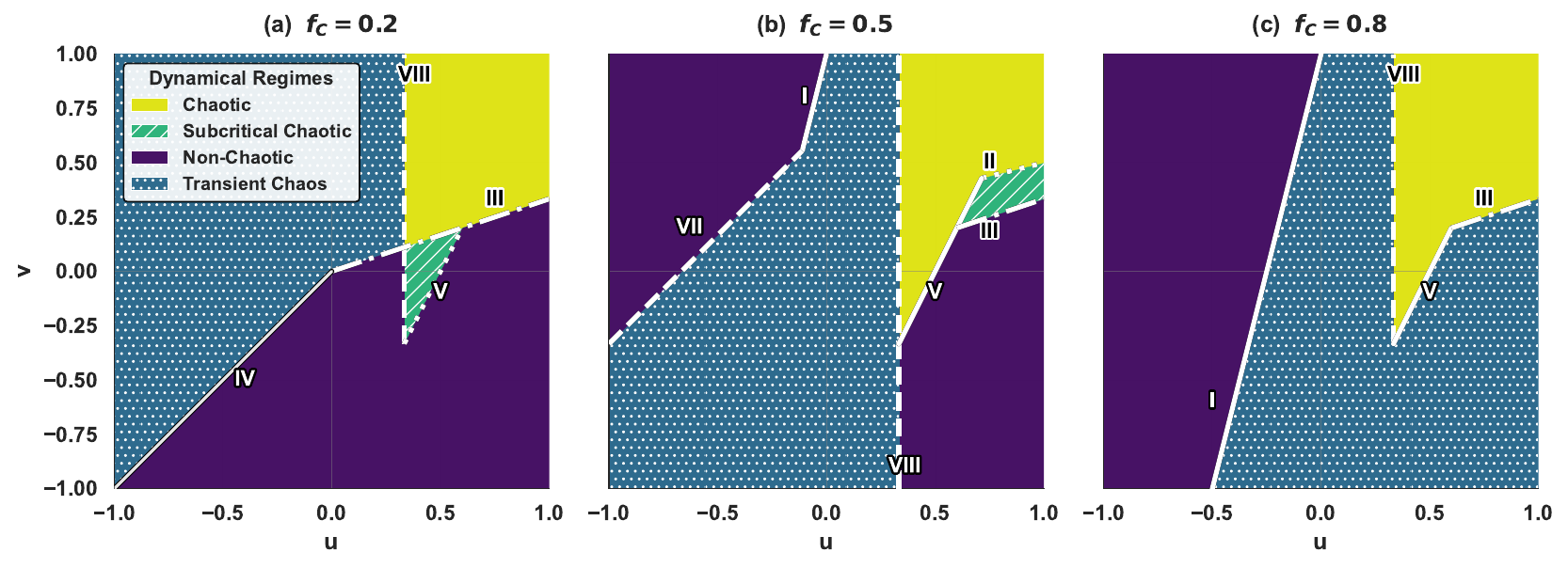}
\caption{\textbf{Three motif instability lines (III, V, VIII)
universally bound the sustained-chaos region across all initial
cooperator densities, indicating the spectral origin of the
phase boundaries.}
Each point in the $(u,v)$ plane is classified by the joint values
of $s$ and $d_\infty$ into one of four dynamical regimes.
\emph{Color key:} dark blue (ordered, $s<1$,
$d_\infty\approx0$); dotted light blue (transient chaos, $s>1$,
$d_\infty\approx0$); yellow (sustained chaos, $s>1$, $d_\infty>0$);
dashed-green (subcritical chaos, $s<1$, $d_\infty>0$).
Dashed lines are the analytical instability lines from
Table~\ref{tab:final_unified}; the boundaries between colored regions
coincide with Lines~III, V, and~VIII across all three panels.
Panels show $f_C(0)=0.2$ (a), $0.5$ (b), $0.8$ (c).}
\label{fig:chaotic_boundaries}
\end{figure*}

Figure~\ref{fig:hamming} overlays the Derrida boundary ($s\approx1$,
yellow contours) on the $d_\infty$ map, and
Fig.~\ref{fig:chaotic_boundaries} shows the resulting four-region
cartography.
Comparing the two contours reveals a systematic discrepancy:
large regions where $s>1$ eventually freeze ($d_\infty\approx0$),
identifying a regime of \emph{transient chaos}.
In the transient chaos regime, the initial
configuration contains locally unstable motifs---so infinitesimal
perturbations grow and $s>1$---but the game's global attractor
eventually organizes the lattice into stable geometric domains,
suppressing all residual damage.
A concrete example is the HG region at $f_C(0)=0.2$: the
predominantly defector background generates motifs with $s>1$, yet cooperation strictly dominates eventually, so the system inevitably converges to the all-$C$ fixed point and freezes.

The same three lines bound the sustained-chaos boundary for all values of $f_C(0)$ which is a nontrivial feature of the cartography. Sustained chaos is an \emph{asymptotic} property ($d_\infty>0$): regardless of the initial density, the long-time dynamics generates a common repertoire of local motifs---cooperative pairs, stripe kinks, and cluster cores---through the continual formation and break-up of clusters, so the boundary is controlled by the same motif-instability lines (III, V, VIII) at every density.
Notably, cooperative cores are essentially absent in the initial
low-$f_C(0)$ lattice yet form dynamically as clusters nucleate and grow, which is why Line~VIII (core erosion) bounds sustained chaos even at $f_C(0)=0.2$.
The boundaries of \emph{sustained} chaos
($d_\infty>0$) are consistent across all three initial conditions and
coincide with specific lines from Table~\ref{tab:final_unified}:
\begin{itemize}
    \item \textbf{Nucleation limit (Line~III, $v=u/3$):}
    For $u>0.6$, sustained activity requires the pair-nucleation
    threshold to be exceeded; below this line cooperative clusters
    cannot self-sustain and the dynamics freezes.
    \item \textbf{Stripe unzipping limit (Line~V, $v=2u-1$):}
    In the SD and PD regions, sustained chaos is confined above
    $v=2u-1$ for $u\in(1/3,3/5)$; below this line, domain walls
    lock into stable stripe patterns and damage cannot propagate
    indefinitely.
    \item \textbf{Core erosion limit (Line~VIII, $u=1/3$):}
    The vertical line $u=1/3$ marks a universal transition between
    transient and sustained chaos across all three initial cooperator
    frequencies, reflecting the purely $u$-driven nature of
    cluster-core instability.
\end{itemize}

Panels (a) and (b) of Fig.~\ref{fig:hamming} further reveal a
\emph{subcritical chaotic phase}: regions where $s<1$ yet
$d_\infty>0$.
Specifically, the triangular region bounded by Lines~III, V,
and~VIII in panel~(a), and the region bounded by Lines~II, III,
and~V in panel~(b), exhibit sustained chaos that the Derrida slope
alone cannot predict.
The mechanism is nonlinear: infinitesimal
perturbations decay ($s<1$), so the linearized framework predicts
stability, but the basin of attraction of the ordered state is
bounded. A finite-amplitude perturbation---one large enough to push
the system across the basin boundary---can reach a self-sustaining
chaotic attractor that the linear analysis cannot see.
Simulations on lattices up to $L=200$ and runs extended to
$t=10^5$ time steps indicate that $d_\infty>0$ in these regions is
not a finite-size or finite-time artifact.
Figure~\ref{fig:chaotic_boundaries} shows the complete cartography.

This subcritical regime parallels the subcritical transition to
turbulence in shear flows, where the laminar base state is linearly
stable yet finite-amplitude perturbations can sustain turbulence~\cite{Avila2011}.
In both settings the ordered (laminar) state is linearly stable while
finite-amplitude perturbations sustain chaotic (turbulent) activity. We
draw the analogy at the level of this shared phenomenology; a
quantitative characterization of the basin boundary and any critical
perturbation amplitude is left to future work.

\section{Conclusion}
\label{sec:conclusio}

This paper asked a simple question: \emph{where do the chaotic phase
boundaries in spatial evolutionary games lie, and why?} The answer is
that each boundary marks the loss of marginal stability of a
characteristic spatial motif. Each threshold is located by a
payoff-balance condition at the motif's contested interface, and
interpreted spectrally as the loss of stability of the corresponding
ordered background.
The framework derives each of these boundaries from a single
microscopic instability mechanism. The boundaries are sharp and
approximately linear, and their positions shift systematically with the
initial cooperator density (Sec.~\ref{qc})---features that were
previously only observed numerically. This supplies the analytical
account of the spatial chaos first demonstrated by Nowak and
May~\cite{nowakmay}.

Our central result is that motif marginal stability sets the chaotic
phase boundaries. Applying an effective-medium argument---the random lattice treated as a mosaic of locally ordered patches---the onset of damage spreading is
controlled by the first statistically prevalent local motif to reach
its threshold ($\Lambda_M=1$). This is what places the analytical lines on top of the
numerical chaos boundaries.
The homogeneous identity $s=\Lambda$ anchors the
spectral interpretation: in this exactly solvable case, the Derrida
slope and the spectral radius of the Boolean Jacobian coincide.
This correspondence addresses the explanatory gap
identified in the introduction. It accounts not only for \emph{that} chaos occurs, but for the \emph{family of motifs} that can trigger it, \emph{where} in parameter
space the transition happens, and \emph{why} the dominant motif shifts
with the initial cooperator density. It also supplies the
mechanism behind a previously unmade connection: the cluster-invasion
thresholds of Hauert~\cite{hauert2001fundamental} and Szab\'{o} and
F\'{a}th~\cite{szabo2007evolutionary} coincide with the payoff-balance
conditions of specific motifs at marginal stability.
The central outcome is not the recovery of any
individual boundary. It is the organization of the entire dynamical
phase diagram by the motif-instability lines of
Table~\ref{tab:final_unified}. The linear-analysis lines set the onset
of chaos ($s\approx1$). A small, fixed subset (notably III, V,
and~VIII) then bounds the \emph{sustained}-chaos region across all
densities studied. The same few lines govern sustained chaos
independently of cooperator density.

We further identify a subcritical chaotic regime ($s<1$, $d_\infty>0$),
in which finite-amplitude perturbations sustain chaos despite linear
stability. This is the spatial evolutionary game analogue of the
subcritical transition to turbulence in shear
flows~\cite{Avila2011}.
Within this regime the ordered state is
\emph{metastable}: robust to infinitesimal perturbations but vulnerable
to sufficiently large structural fluctuations.

Several natural extensions remain open.
First, asynchronous updating and stochastic imitation rules break the
synchronous, deterministic one-step structure on which the
block-circulant form of the Boolean Jacobian relies. Stochastic noise
is then expected to smear the sharp spectral boundaries into crossover
regions~\cite{Amaral2016}.
Second, extending the framework to heterogeneous topologies requires
replacing the DFT with the graph Fourier transform, whose eigenbasis
is set by the graph Laplacian rather than by translational symmetry.
Third, conformity dynamics create effective surface tension and smooth
cooperative boundaries~\cite{Szolnoki2015}. These may interact
nontrivially with the motif instabilities identified here, potentially
shifting or splitting the critical lines.
Each extension preserves the core logic of the
framework---linearize, decompose, find the marginal-stability
condition---while replacing the specific technical ingredients. The
present results therefore serve as the analytically tractable baseline to which these additional effects can be added systematically.

Beyond evolutionary games, the ingredients of this analysis are
generic. The exact identity $s=\Lambda$ for a homogeneous background,
and the motif-threshold equivalence for disordered ones, require only a
deterministic, translationally invariant update rule with Boolean
(idempotent) switching coefficients. The linearized operator is then
block-circulant on each ordered patch, and the scalar idempotency
argument applies locally~\cite{bagnoli1992damage,vispoel2024damage}.
None of this depends on the payoff interpretation: what matters is the local block-circulant structure of the linearized operator and the
idempotency of the Boolean coefficients. The link between damage
spreading and spectral instability is therefore a structural feature of
this class of cellular automata, not specific to evolutionary games.

\section*{Data Availability}
The data, code, and materials supporting the findings of this article 
are available at \url{https://aydogmusozgur.github.io/code-data/spectral_cartography.html}
\begin{acknowledgments}
The author thanks the two anonymous referees for their careful reading
and constructive comments, which substantially improved the manuscript.
An AI language model was used for limited language editing of an earlier
draft; the present version was revised by the author, who takes full
responsibility for the scientific content.
\end{acknowledgments}

\appendix

\section{The Derrida--Spectral Correspondence}
\label{app:spectral_proof}

This appendix collects the technical material supporting
Sec.~\ref{subsec:formalism}: the derivation of the exact
homogeneous identity $s=\Lambda$ (Sec.~\ref{app:idempotency}), the Bloch structure of the Boolean Jacobian for
periodic backgrounds (Sec.~\ref{app:bloch}), and the
effective-medium threshold argument for random configurations
(Sec.~\ref{app:threshold}).

Throughout we use one structural fact. For a homogeneous reference
background, the Boolean Jacobian is translationally invariant and therefore
block-circulant; it is diagonalized exactly by the discrete Fourier
transform~\cite{davis1979circulant}. For a periodic background, the Jacobian
inherits the periodicity of the unit cell and admits a Bloch--Fourier
decomposition~\cite{cross1993}, leading to a finite Bloch symbol
$\widehat{J}(\mathbf{k})$. For a random configuration this structure holds
only locally, motivating the effective-medium argument of
Sec.~\ref{app:threshold}.

\subsection{The exact identity $s=\Lambda$ for homogeneous backgrounds}
\label{app:idempotency}

A spatially random perturbation with flat power spectrum excites all Fourier
modes equally. The one-step growth of the total damage---measured by the
squared norm $\|\boldsymbol{\delta}\|^2$, which equals
the number of damaged sites for Boolean $\delta$---is therefore an average over all modes,
weighted by their squared amplification factors $|\lambda(\mathbf{k})|^2$.
For a general linear operator this mode average and the spectral radius
$\Lambda=\max_{\mathbf{k}}|\lambda(\mathbf{k})|$ are unrelated quantities.
For the homogeneous Boolean Jacobian they coincide, forced together by the
idempotency of the coefficients, $J_0^2=J_0$ and $J_*^2=J_*$.

Expand $|\lambda(\mathbf{k})|^2=(J_0+2J_*[\cos k_x+\cos k_y])^2$ and average
over the Brillouin zone. The cross terms between $J_0$ and the cosines vanish
by symmetry, as does the mixed term
$\langle\cos k_x\cos k_y\rangle=0$; each cosine-squared term averages to
$\langle\cos^2 k_x\rangle=\tfrac12$. The zone average is therefore
\begin{equation}
\big\langle|\lambda(\mathbf{k})|^2\big\rangle
= J_0^2+4J_*^2
= J_0+4J_*,
\label{eq:zone_average}
\end{equation}
where the second equality applies idempotency. Since $J_0,J_*\ge0$, the
maximum of $|\lambda(\mathbf{k})|$ is attained at the uniform mode
$\mathbf{k}=(0,0)$, where $\cos k_x+\cos k_y$ peaks, and equals
$J_0+4J_*$ as well. Because Boolean damage satisfies
$\|\boldsymbol{\delta}\|^2=L^2 d$, the zone-averaged mean-square
amplification (Eq.~\eqref{eq:zone_average}) is precisely the Derrida slope, and
\begin{equation}
s=\Lambda=J_0+4J_*.
\end{equation}
This establishes Eq.~\eqref{eq:derrida_spectral} of the main text. The
argument uses only translational invariance (so that plane waves diagonalize
the Jacobian) and idempotency of the Boolean coefficients; it is therefore a
structural property of this class of cellular automata rather than a feature
of the game-theoretic payoffs.
\subsection{Bloch structure for periodic backgrounds}
\label{app:bloch}

For a background with a nontrivial unit cell---an alternating stripe or a
cooperative core---the payoff ordering varies between site classes within the
cell, so the Boolean derivative is no longer the same at every site. A
cooperator in the interior of a cluster ($n_C=4$) has a different switching
sensitivity than one at the cluster boundary ($n_C=2$ or $3$), and the
coefficient for a north neighbor may differ from that for an east neighbor
when the background breaks the four-fold rotational symmetry. The linearized
rule therefore carries one coefficient set per site class,
\begin{equation}
\delta_{n,m}(t+1)=
\sum_{(r,s)\in\mathcal{N}_0}
J^{(\alpha(n,m))}_{r,s}\,\delta_{n+r,m+s}(t),
\label{eq:linearized_bloch}
\end{equation}
where $\alpha(n,m)$ identifies the site class at $(n,m)$. The coefficients
are not invariant under single-site translations, but they are invariant
under translations by a full unit cell (e.g.\ by two sites for an alternating
stripe). Grouping sites into unit cells, the Jacobian is circulant at the
level of cells, with each entry a small matrix acting on the site-class
indices---that is, block-circulant. It is diagonalized by a Bloch--Fourier
transform, with eigenvalues given by those of a finite Bloch symbol
$\widehat{J}(\mathbf{k})$ whose dimension equals the number of site classes,
and the marginal-stability criterion is
$\rho(\widehat{J}(\mathbf{k}))=1$.

Because the Boolean coefficients are piecewise constant in $(u,v)$ and change
only when a local payoff ordering flips, the Bloch symbol is likewise
piecewise constant within each payoff-ordering region. The loci at which its
stability properties can change are therefore exactly the payoff-balance
lines of the corresponding motifs. This is why, in the main text, the
instability thresholds are obtained directly from payoff balance at each
motif's contested interface (Sec.~\ref{subsec:derivations}) rather than by
explicit diagonalization of $\widehat{J}(\mathbf{k})$; the resulting
boundaries are verified against the numerical Derrida contours in
Fig.~\ref{fig:chaotic_boundary_lines}.

We note that for these multi-class symbols the scalar idempotency argument of
Sec.~\ref{app:idempotency} does not apply, and the identity $s=\Lambda$ is
not expected to hold exactly; the periodic motifs are treated at the
threshold level only.

\subsection{Threshold equivalence for random configurations}
\label{app:threshold}

When the reference configuration is drawn randomly at density $f_C$, the
Boolean Jacobian $\mathbf{J}(S^0)$ depends on the realization $S^0$ and is no
longer globally block-circulant, so the equality $s=\Lambda$ cannot be
expected to hold pointwise. We establish the threshold
equivalence~\eqref{eq:threshold} through an effective-medium argument, valid
in the thermodynamic limit, where
$\Lambda_M=\max_{\mathbf{k}}\rho(\widehat{J}_M(\mathbf{k}))$ is the spectral
radius of the (possibly multi-band) Bloch symbol of motif $M$
(Sec.~\ref{app:bloch}).

\paragraph{Heuristic physical motivation.}
The equivalence~\eqref{eq:threshold} is not a rigorous mathematical proof but
an effective-medium estimate. A random configuration can be viewed as a
spatial superposition of local geometries (single invaders, pairs, stripes,
kinks, cores) whose relative frequencies are determined by $f_C(0)$. Each
geometry possesses an associated local instability threshold $\Lambda_M$.
When all statistically prevalent motifs are stable ($\Lambda_M<1$),
perturbations fail to grow in the environments that dominate the lattice, and
the ensemble-averaged damage contracts ($s<1$). Conversely, once a
statistically prevalent motif becomes unstable ($\Lambda_M>1$), perturbations
are linearly amplified in a finite fraction of local environments across the
lattice, leading to macroscopic growth of the ensemble-averaged damage
($s>1$).

\paragraph{Consequence.}
Within this picture, the transition $s=1$ occurs when the first statistically
prevalent local geometry loses marginal
stability~\cite{bagnoli1992damage}, which explains the close agreement
between the analytically derived boundaries
(Table~\ref{tab:final_unified}) and the numerically measured Derrida contours
(Fig.~\ref{fig:chaotic_boundary_lines}), even for maximally disordered
initial conditions ($f_C(0)=0.5$).

\paragraph{Remark on the magnitude of $s$ in the chaotic interior.}
Deep inside the chaotic phase, the numerical Derrida slope and the spectral
radius of any single ordered motif may differ quantitatively, because the
random Jacobian mixes contributions from multiple local geometries with
distinct instability spectra. The threshold
correspondence~\eqref{eq:threshold} depends only on the \emph{sign} of the
linear growth rate and is insensitive to such interior deviations.

\paragraph{Scope.}
The arguments above address only the linear regime ($d_0\to0$). Chaotic
dynamics may persist even where $s<1$, driven by finite-amplitude
perturbations rather than linear instability (Sec.~\ref{sec:hamming}); this
subcritical regime lies outside the scope of the spectral ansatz and is
characterized numerically through $d_\infty$.
\sloppy
\setlength{\emergencystretch}{3em}
\bibliographystyle{apsrev4-2}
\bibliography{references}

\end{document}